\documentclass[useAMS,usenatbib]{mn2e}
\usepackage{aas_macros,graphicx,times,multirow,amsmath}

\title[A 292-s X-ray pulsar in the SMC]{CXOU\,J005047.9--731817: a 292-s X-ray binary pulsar in the Small Magellanic Cloud}
\author[P.~Esposito et al.] {P.~Esposito,$^{1}$\thanks{E-mail: paoloesp@iasf-milano.inaf.it} G.~L.~Israel,$^{2}$ L.~Sidoli,$^{1}$ G.~A.~Rodr\'iguez~Castillo,$^{2,3}$ N.~Masetti,$^{4}$
\newauthor P.~D'Avanzo$^5$ and S.~Campana$^5$\smallskip\\
$^1$Istituto di Astrofisica Spaziale e Fisica Cosmica - Milano, INAF, via E. Bassini 15, I-20133 Milano, Italy\\
$^2$Osservatorio Astronomico di Roma, INAF, via Frascati 33, I-00040 Monteporzio Catone, Italy\\
$^3$Dipartimento di Fisica, Universit\`a di Roma ``La Sapienza", p.le A. Moro 2, I-00185 Roma, Italy\\ 
$^4$Istituto di Astrofisica Spaziale e Fisica Cosmica - Bologna, INAF, via P. Gobetti 101, I-40129 Bologna, Italy\\
$^5$Osservatorio Astronomico di Brera, INAF, via E. Bianchi 46, I-23807, Merate (LC), Italy 
}
\date{Accepted 2013 June 4.  Received 2013 June 4; in original form 2013 May 20} \pagerange{\pageref{firstpage}--\pageref{lastpage}} \pubyear{2013}

\def\LaTeX{L\kern-.36em\raise.3ex\hbox{a}\kern-.15em
    T\kern-.1667em\lower.7ex\hbox{E}\kern-.125emX}

\def\cxo {\emph{Chandra}}

\def\src {CXOU\,J0050}
\def\flux {\mbox{erg cm$^{-2}$ s$^{-1}$}}
\def\lum {\mbox{erg s$^{-1}$}}
\def\nh {$N_{\rm H}$}

\begin{document}

\label{firstpage}
\maketitle
\begin{abstract}
We report on the discovery of a transient X-ray pulsars, located in the Small Magellanic Cloud, with a pulse period of 292 s. A series of \cxo\ pointings fortuitously recorded in 2010 April--May the occurrence of a two-weeks-long outburst, during which the source luminosity increased by a factor of about 100, reaching a peak of $\sim$$10^{36}$ \lum\ (for a distance of 61 kpc). Complex-shape and energy-dependent pulsations were detected close to the outburst peak and during the very first part of its decay phase. During the outburst, the phase-averaged spectrum of the pulsar was well described by an absorbed power law with photon index $\Gamma \sim0.6$, but large variations as a function of phase were present. The source was also detected by \cxo\ several times (during 2002, 2003, 2006, and 2010) at a quiescent level of $\sim$$10^{34}$ \lum. In 2012 we performed an infrared photometric follow-up of the $R\sim15$ mag optical counterpart with the ESO/VLT and a spectroscopic observation by means of the CTIO telescope. The optical spectra suggest a late-Oe or early-Be V--III luminosity-class star, though a more evolved companion cannot be ruled out by our data (we can exclude a luminosity class I and a spectral type later than B2). Finally, we show that the outburst main parameters (duration and peak luminosity) can be accounted for by interpreting the source transient activity as a type I outburst in a Be X-ray binary.          
\end{abstract}

\section{Introduction}
The  Small Magellanic Cloud (SMC) provides a wonderful laboratory for the study of massive X-ray binaries (XRBs), since distances have small relative  uncertainty and the interstellar absorption is low.  Moreover, SMC underwent a recent star formation episode which triggered the formation of a large population of high mass X-ray binaries (HMXBs), especially Be-type binary systems, which are overabundant with respect to the Milky Way \citep{mcbride08}.\\ 
\indent CXOU\,J005047.9--731817 (hereafter \src) is a faint X-ray source discovered during observations performed to survey the HMXB population in the SMC (\citealt{shtykovskiy05,laycock10}; \citealt*{sarraj12}). The sub-arcsec position uncertainty obtained through a long \cxo\ observations made it possible to associate the source to an optical counterpart \citep{laycock10}. This counterpart, a blue early-type candidate ($V=15.07$ mag), suggests that \src\ is a massive X-ray binary. Sinusoidal optical variations with amplitude of 0.017 mag and periodicity of 1.05 days, most likely non-radial pulsations, have been reported for the companion star of \src\ \citep*{sarraj12}.\\
\indent Here we report on the discovery of X-ray pulsations at \mbox{292 s} from \src\ using a series of \cxo\ observations collected in 2010 April--May. We also analysed all other available \cxo\ data and found that in 2010 April--May \src\ was in a particularly bright state that lasted approximately two weeks. We observed the field of \src\ at infrared and optical wavelengths with VLT and CTIO telescopes; this allowed us to obtain some constraints on the nature of its stellar companion.

\section{\emph{Chandra} X-ray observations}
\src\ was observed serendipitously by \cxo\ 22 times in observations targeting the supernova remnant B\,0049--73.6 or surveying the SMC (\citealt{edge04}; \citealt*{hendrick05}; \citealt{guerrero08,antoniou09,antoniou10,laycock10}). Most observations were carried out in 2010 April--June but some date back to 2006, 2003 and 2002. The data were acquired with the Advanced CCD Imaging Spectrometer (ACIS; \citealt{garmire03}) Imaging or Spectroscopic array in Very Faint imaging (Timed Exposure) mode and in full frame, with readout time of 3.14 s. See Table\,\ref{logs} for a summary of the pointings.

All data were reprocessed with the Chandra Interactive Analysis of Observations software (\textsc{ciao}, version 4.4) and the calibration database \textsc{caldb} 4.4.8. Source spectra and event lists were extracted from circles of $\sim$4--5 arcsec radii (depending on the off-axis angle), while for the background we used annuli around the source with internal (external) radius of 15 (30) arcsec. The spectra, the ancillary response files and the spectral redistribution matrices were created using \textsc{specextract}. For the timing analysis, the photon arrival times were transformed to Barycentric Dynamical Time (TDB) using \textsc{axbary} and the coordinates of the source given in Section\,\ref{vltobs}.

\begin{table}
\centering \caption{\cxo\ observations used for this work.} \label{logs}
\begin{tabular}{@{}lccccc}
\hline
Obs.\,ID  & Date & Exposure & Count rate$^{a}$ \\
 & & (ks) & ($10^{-2}$ cts s$^{-1}$) \\
\hline
2945 & 2002~Oct~02 & 11.8 & $0.11\pm0.03^{b}$  \\
3907 & 2003~Feb~28--Mar~01 & 50.8 &  $0.09\pm0.01$  \\
8479,\,7156,\,8481$^{c}$ & 2006~Nov~21--23 & 98.0 & $0.043\pm0.007^{b}$  \\ 
11097 & 2010~Apr~14--15 & 30.0 & $0.17\pm0.08$  \\
11980 & 2010~Apr~21 & 23.1 & $0.13\pm0.02$  \\
12200 & 2010~Apr~22--23 & 27.1 & $0.30\pm0.03$  \\
11981 & 2010~May~01--02 & 34.0 & $9.15\pm0.17$ \\
12208 & 2010~May~02 & 16.2 & $9.88\pm0.25$ \\
11982 & 2010~May~03 & 24.9 & $10.0\pm0.2\phantom{0}$  \\
12211 & 2010~May~04 & 35.5 & $9.64\pm0.17$ \\
12210 & 2010~May~07--08 & 27.9 & $6.30\pm0.15$ \\
11983 & 2010~May~12 & 29.7 & $3.1\pm0.1$ \\
12212 & 2010~May~14--15 & 18.5 & $1.9\pm0.1$  \\
11095 & 2010~May~18 & 19.1 & $0.32\pm0.04$  \\
12215 & 2010~May~19 & 23.0 & $0.17\pm0.03$  \\
11984 & 2010~May~22 & 20.8 & $0.19\pm0.03$  \\
12216 & 2010~May~22--23 & 29.1 & $0.20\pm0.03$ \\
11985 & 2010~May~25 & 21.0 & $0.34\pm0.04$  \\
12217 & 2010~May~28 & 29.1 &  $0.12\pm0.02$  \\
11096 & 2010~Jun~02 & 37.1 & $0.17\pm0.02$  \\
\hline
\end{tabular}
\begin{list}{}{}
\item[$^{a}$] 0.3--10 keV  net count rate, not corrected for PSF and vignetting effects.
\item[$^{b}$] ACIS-I data.
\item[$^{c}$] These observations were combined to increase the count statistics.
\end{list}
\end{table}

\subsection{Timing analysis and results}\label{timing}

\begin{figure}
\centering
\resizebox{\hsize}{!}{\includegraphics[angle=-90]{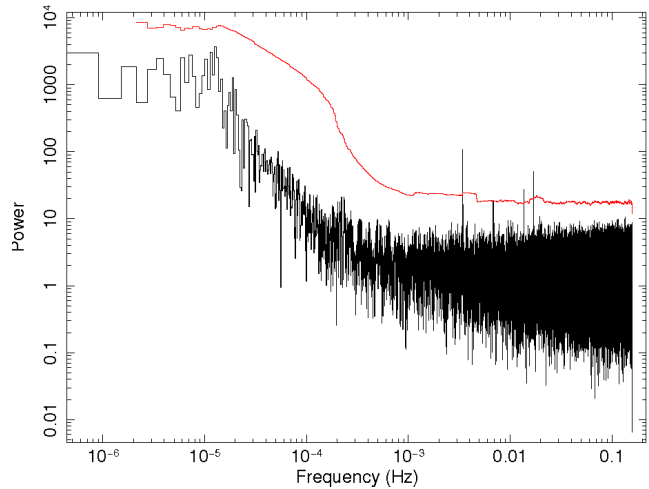}}
\caption{\label{powerspec} Fourier power spectrum computed with the whole \cxo\ data-set of \src. The red stepped line corresponds to 3.5$\sigma$ confidence level threshold for potential signals (estimated taking into account the number of trials, equal to the number of frequency bins of the spectrum).}
\end{figure}

Pulsed emission at $\sim$292 s was discovered from \src\ within the \emph{Chandra ACIS Timing Survey at Brera And Rome astronomical observatories} (CATS@BAR) project (\citealt{eis13}; Israel et al., in preparation). CATS@BAR is a Fourier-transform-based systematic search for new pulsating sources in the \cxo\ ACIS public archive. Fig.\,\ref{powerspec} shows the Fourier periodogram (computed using using all the 2010 \cxo\ data, see Table\,\ref{logs}) that led to the discovery of pulsations in \src. A peak at frequency $\nu=1/P \simeq 3.415$ mHz ($P\simeq292$ s) stands well above the 3.5$\sigma$ significance level, which was estimated following \citet{israel96}. The inspection of the periodogram revealed also several higher harmonics of the coherent pulsation, $\nu_n=1/(nP)$, with $n = 2$, 4, and 5.
\begin{figure}
\resizebox{0.85\hsize}{!}{\includegraphics[angle=0]{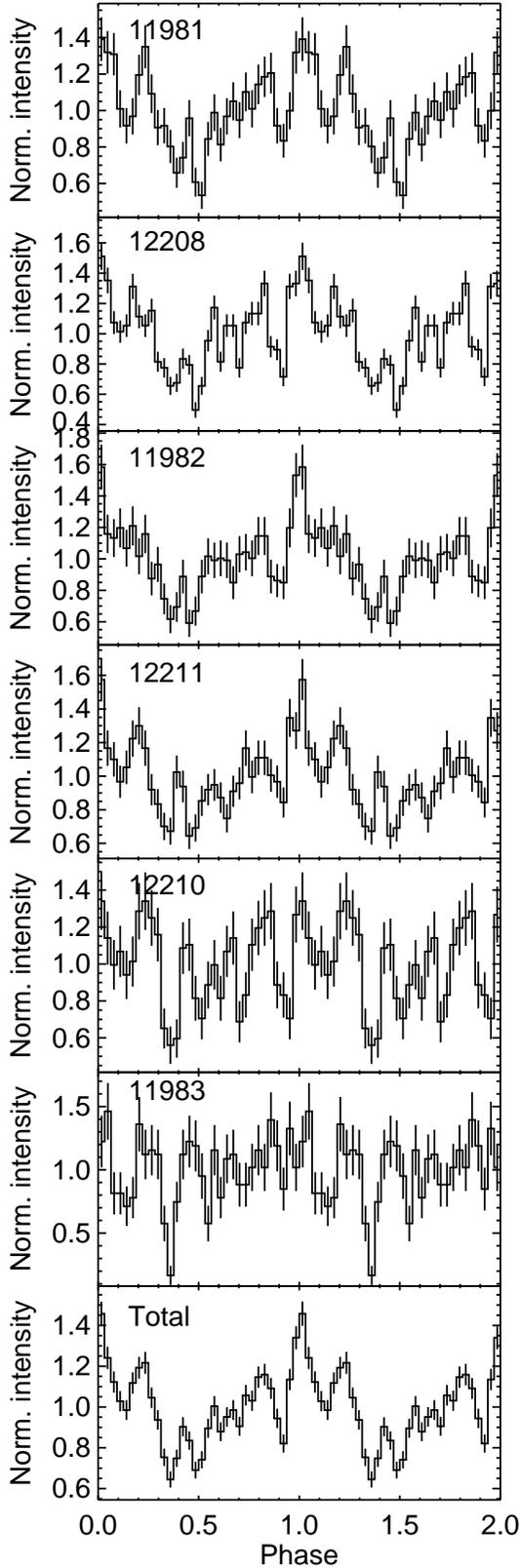}}
\caption{\label{profiles} Background-subtracted epoch-folded \cxo\ 0.3--10 keV pulse profiles (we assumed the $P$--$\dot{P}$ solution of Sect.\,\ref{timing}; the TDB  reference epoch of the maximum is MJD 55320.0034(1)).}
\end{figure}
\begin{figure}
\centering
\resizebox{0.85\hsize}{!}{\includegraphics[angle=0]{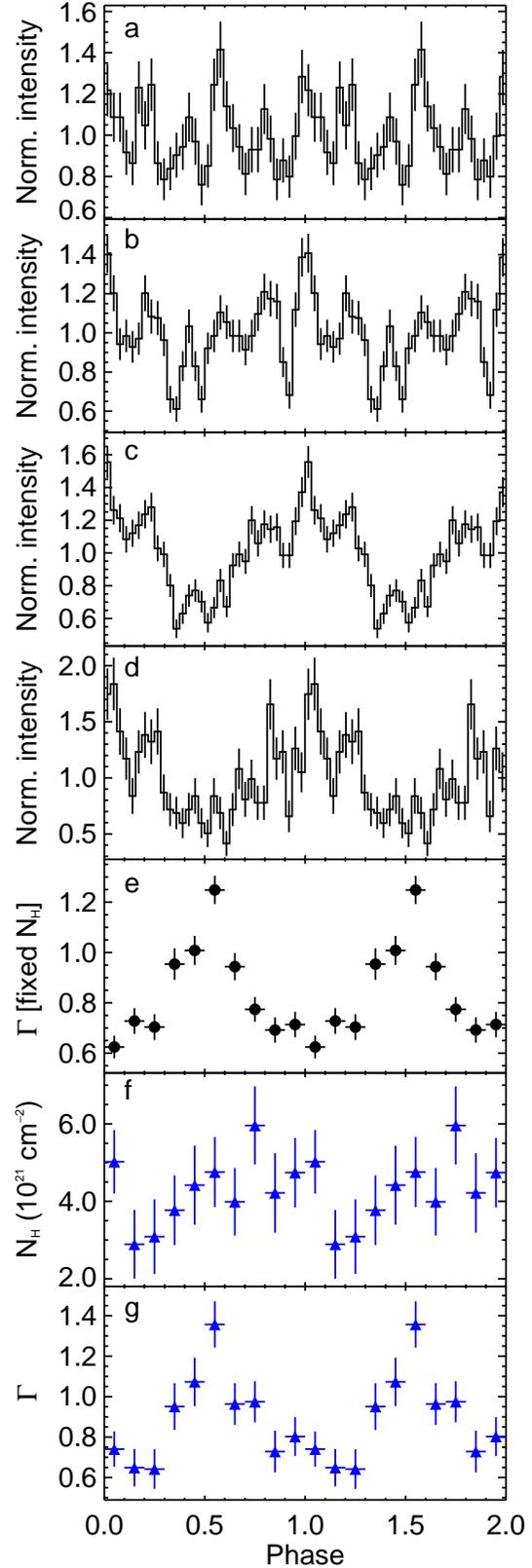}}
\caption{\label{4band}  Background-subtracted epoch-folded total \cxo\ pulse profiles in 0.3--1.5 keV (a), 1.5--3 keV (b), 3--6 keV (c), and 6--10 keV (d). The bottom panels show the photon index and absorption variations across the pulse phase. In panel (e) the \nh\ was held fixed to the average value; panels (f) and (g) correspond to the fit in which both \nh\ and photon index were left free to vary (see text).}
\end{figure}
\begin{table}
\centering \caption{\cxo\ timing results.} \label{periods}
\begin{tabular}{@{}lcc}
\hline
Obs.\,ID  & Period$^{a}$ & Pulsed fraction$^{b}$\\
 & (s) & (\%)\\
\hline
11980,12200$^c$ & -- & $<$65 \\
11981 & $292.705\pm0.048$ & $23.5\pm3.3$\\
12208 & $292.78\pm0.15$ & $27.5\pm4.2$\\
11982 & $292.642\pm0.083$ & $22.0\pm3.4$ \\
12211 & $292.793\pm0.049$ & $20.4\pm3.2$\\
12210 & $292.771\pm0.057$ & $22.0\pm4.3$ \\
11983 & $292.624\pm0.071$& $35.7\pm4.7$ \\
12212 & -- & $<$65 \\
11095,12215$^c$ & -- & $<$80 \\
11985,12217,11096$^c$ & -- & $<$70 \\
\hline
\end{tabular}
\begin{list}{}{}
\item[$^{a}$] Periods were derived from a $Z^2_5$ test; uncertainties in the last digits were determined from Monte Carlo simulations similar to those described in \citet*{gotthelf99}.
\item[$^{b}$] When no periodicity was found, a 3$\sigma$ upper limit (computed following \citealt{vaughan94short}) is given instead of a value and associated error.
\item[$^{c}$] These observations were combined to increase the count statistics.
\end{list}
\end{table}

We performed a period search around 292 s in all the individual observations and in combined data from few contiguous pointings using a $Z^2_n$ test (e.g. \citealt{buccheri83short}). The signal can be detected with good confidence (always above $4.5\sigma$) in the observations performed from 2010 May 01 (Obs.\,ID 11981) to May 12 (Obs.\,ID 11983; see Table\,\ref{periods} and Fig.\,\ref{outburst}). The pulsed fraction of the folded profiles, which we define as $(M-m)/(2(M+m))$, where $M$ and $m$ are the maximum and minimum background-subtracted count rates in the pulse profile, respectively, varies from $\sim$20\% to 35\%. The measured period and pulsed fraction values are given in Table\,\ref{periods}.

In the other observations no significant pulsation was detected. We note that with the exception of the 2010 May 14 observation (Obs.\,ID 12212) which has $\approx$350 source photons, all other observations have less than 100 net counts so, also considering the moderate pulsed fraction, the nondetections are not surprising. In Table\,\ref{periods} we report the (poorly constrained) upper limits on the pulsed fraction obtained for observation 12212 and a few combined temporally-nearby observations in which at least 150 source counts were present.

A linear, unweighted fit of the periods in Table\,\ref{periods} suggests a period derivative $\dot{P}\approx-9\times10^{-8}$ s s$^{-1}$, but with a rather large uncertainty. In order to obtain a more precise assessment, beginning from the four closely-spaced observations collected in 2010 May 01--04 we started to build coherent rotational ephemeris linking all data until May 12 (Obs.\,ID 12212). While the data are consistent ($\chi^2_\nu=1.94$ for 4 degrees of freedom (dof)) with a constant period of $292.7836\pm0.0033$ s (here and throughout the paper, uncertainties are given at 1$\sigma$ confidence level), this solution results in a large scatter of the pulse phases in the folded profile. A much better phasing can be obtained with the introduction of a $\dot{P}$ component ($\chi^2_\nu=1.06$ for 3 dof). The significance of the period derivative, as estimated by the Fisher test, is 2.36$\sigma$. The parameters of this timing solution, valid over   MJD 55317--55328, are $P_0=292.7796\pm0.0054$ s and $\dot{P} = (-9.6\pm2.1)\times10^{-8}$ s s$^{-1}$; the TDB reference epoch for the period and its derivative is $T_0 =$~MJD 55320.000. This is a rather intense spin-up rate, but not unusual among accreting X-ray pulsar. For example, a spin-up rate of $\sim$$-1.5\times10^{-7}$ s s$^{-1}$ was measured in the transient Be HMXB pulsar (spin period of $\sim$358 s) SAX\,J2103.5+4545 \citep{sidoli05}. We also note that, lacking information on the orbit of \src, it is not possible to estimate the contribution of the orbital motion to the observed spin-period derivative.

In Fig.\ref{profiles} we show the data folded using the $P$--$\dot{P}$, as well as the total pulse profile obtained by summing coherently all data between 2010 May 01 and 12. It is apparent that the pulse shape is complex and evolved through the observations (and flux, see Section\,\ref{spectralanalysis}). Also, the profile is strongly energy dependent. Fig.\,\ref{4band} shows the summed profile in four energy bands (0.3--1.5 keV, 1.5--3 keV, 3--6 keV, and 6--10 keV). In the soft band its shape is rather complex, with four peaks, while at intermediate energies the profile is more sinusoidal, but several substructures are observable, superimposed on a broad wave.

\subsection{Spectral analysis and results}\label{spectralanalysis}
For the spectral analysis of \src, we started from the seven spectra with more than 300 source counts, namely those collected from 2010 May 01 (Obs.\,ID 11981) to May 15 (Obs.\,ID 12212). The spectra were grouped so as to have a minimum of 20 counts per energy bin and the spectral fitting was performed with the \textsc{xspec} package (version 12.7; \citealt{arnaud96}) in the range 0.3--8 keV.

A simple power-law model, corrected for the interstellar absorption, provides a satisfactory fit for all data-sets. The results are summarised in Table\,\ref{fits}. The fit equivalent absorbing column is $N_{\rm H} \simeq4\times10^{21}$ cm$^{-2}$, a value similar to the total Galactic H\textsc{i} column density in the direction to the source \citep{kalberla05}. The power-law photon index is rather flat ($\Gamma\sim0.8$) and the flux decreased from $\sim$(2.2--$2.5)\times10^{-12}$ \flux\ during the first observations to $\sim$$5\times10^{-13}$ \flux\ on 2010 May 14--15. Under the assumption that \src\ is located in the SMC (distance 61 kpc; \citealt*{hilditch05,keller06}), the maximum observed luminosity was $\sim$$1.2\times10^{36}$ \lum\ on May 03 (Obs.\,ID 11982).
\begin{table*}
\centering 
\begin{minipage}{12.1cm}
\caption{Spectral analysis of \src. Errors are at a 1$\sigma$ confidence level for a single parameter of interest.} \label{fits}
\begin{tabular}{@{}cccccc}
\hline
Observation  & \nh & $\Gamma$ & Flux$^a$ & Luminosity$^a$ & $\chi^2_\nu$ (dof)\\
& ($10^{21}$ cm$^{-2}$) & & ($10^{-12}$ \flux) & ($10^{36}$ \lum) &\\
\hline
11981 & $4.6\pm0.5$ & $0.91\pm0.06$ & $2.22\pm0.07$ & $1.05\pm0.03$ & 1.06 (119) \\
12208 & $4.0\pm0.7$ & $0.91\pm0.09$ & $2.35\pm0.11$ & $1.11\pm0.04$ &0.92 (66) \\
11982 & $4.2\pm0.6$ & $0.85\pm0.07$ & $2.51\pm0.09$ & $1.18\pm0.04$ & 0.99 (100) \\
12211 & $3.5^{+0.4}_{-0.5}$ & $0.76\pm0.06$ & $2.46\pm0.08$ & $1.14\pm0.03$ & 1.08 (128) \\
12210 & $3.8^{+0.8}_{-0.7}$ & $0.79^{+0.09}_{-0.08}$ & $1.65\pm0.07$ & $0.77\pm0.03$ & 0.73 (73) \\
11983 & $5\pm1$ & $0.98\pm0.13$ & $0.72\pm0.05$ & $0.35\pm0.02$ & 1.52 (39) \\ 
12212 & $<11^{b}$ & $0.7\pm0.2$ & $0.54\pm0.05$ & $0.25\pm0.02$ & 0.40 (14) \\
\cline{0-0}
average & $3.8\pm0.3$ &  $0.83^{+0.03}_{-0.04}$ & -- & -- & 0.98 (430) \\
\hline
\end{tabular}
\begin{list}{}{}
\item[$^{a}$] In the 1--10 keV energy range; for the luminosity we assumed a distance of 61 kpc.
\item[$^{b}$]Upper limit (3$\sigma$ c.l.).
\end{list}
\end{minipage}
\end{table*}

\begin{figure*}
\centering
\resizebox{.7\hsize}{!}{\includegraphics[angle=0]{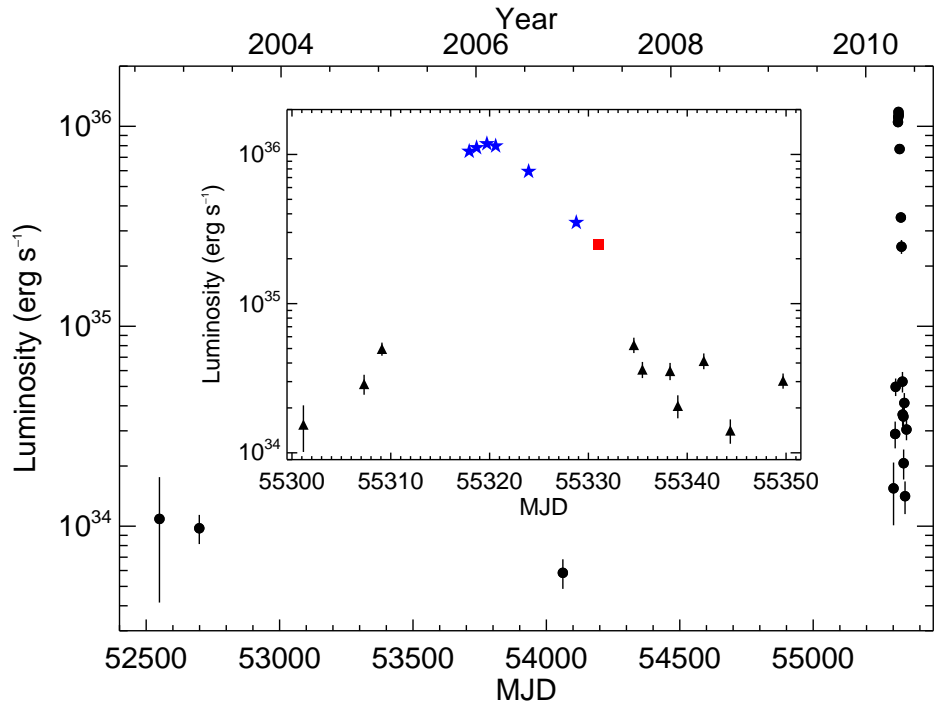}}
\caption{\label{outburst} Long-term light curve (1--10 keV luminosity, assuming a distance of 61 kpc). The inset shows in detail the luminosity evolution during the 2010 outburst. The blue stars and the red square mark the observations with enough source counts to provide high-statistics spectra; in particular, the blue stars are the observations in which the 292-s pulsations were detected, while the red square indicates Obs.~12212, in which no pulsations were observed. The triangles indicate the observations for which the flux was derived from the count rate by assuming the average spectral parameters of Table\,\ref{fits} (see text).}
\end{figure*}

All other observations have too few source counts to allow for a meaningful spectral analysis. For this reason, using \textsc{xspec} and the appropriate response matrices, we converted the count rates into luminosity adopting the average \nh\ and photon index given in Table\,\ref{fits}. The resulting long-term light curve is shown in Fig.\,\ref{outburst}. It suggests that the typical quiescent luminosity of \src\ is $\approx$(5--$10)\times10^{34}$ \lum\ and that in 2010 May the source underwent an outburst that lasted roughly 2 weeks.

Since the energy-resolved pulse profiles of \src\ (Fig.\,\ref{4band}) indicate strong spectral variability across the pulse cycle, we performed a phase-resolved spectral analysis by means of the six observations in which the 292-s pulsations is detected (Section\,\ref{timing}). The data were phase-binned using the \textsc{ciao} task \textsc{dmtcalc} and using the $P$--$\dot{P}$ solution derived with our timing analysis (Section\,\ref{timing}). Then, we extracted for each observation 10 spectra from intervals of equal phase width (0.1 cycles) and combined the spectra from the same phase bins using \textsc{combine\_spectra}, which also averages the response matrices. This resulted in ten spectra with 16.6-ks exposure each and a number of net source counts ranging from $\sim$1000 to 1700. We fit to all spectra simultaneously an absorbed power-law model with the \nh\ fixed to the average value, $3.8\times10^{21}$ cm$^{-2}$ (Table\,\ref{fits}), and all other parameters left free. This yielded a good fit ($\chi^2_\nu=0.93$ for 565 dof) that shows that the photon index was modulated at the 292-s period, describing a `shark-tooth' profile across the pulse cycle (Fig.\,\ref{4band}). The spectrum is softer around the absolute pulse maximum ($\Gamma_{\mathrm{S}} = 1.25 \pm 0.06$) and harder in the pulse minimum  ($\Gamma_{\mathrm{H}} = 0.62 \pm 0.05$). The results were essentially identical when the absorbing column was left free to assume a different value for each spectrum. A constant fit of the \nh\ values yields a $\chi^2_\nu=8.75$ for 9 dof, indicating little or no measurable variability (see Fig.\,\ref{4band}). Both the fine structures of the pulse profiles and the fit residuals in some phase-resolved spectra hint at the presence of narrow and phase-dependent spectral features (such as absorption/emission lines) but the relatively low count statistics of the data did not allow us to investigate this in detail. 

\section{Infrared Observations}\label{vltobs}
\begin{figure}
\resizebox{\hsize}{!}{\includegraphics[angle=0]{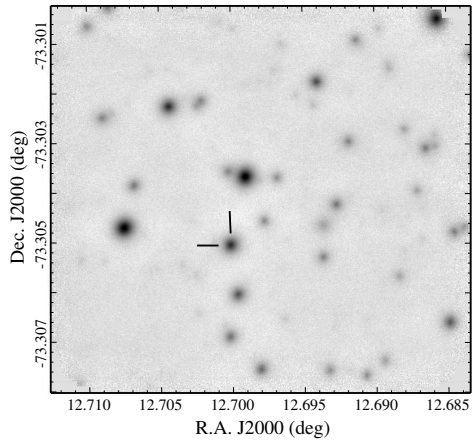}}
\caption{\label{findchart} VLT $K_{\rm S}$-band image (30-arcsec side) of the field of \src. The infrared countepart of the pulsar is marked with solid lines.}
\end{figure}

The observations in $K_{\rm S}$ band presented here were performed in 2012 July at the European Southern Observatory (ESO) Very Large Telescope (VLT) 8.2-m Unit 4 (Yepun) Telescope at Cerro Paranal and the data were acquired with the Nasmyth Adaptive Optics System and the High Resolution Near Infrared Camera (NAOS-CONICA, or NaCo; \citealt{rousset03short,lenzen03short}) providing a pixel size of 0\farcs027.  We observed the field of \src\ in 20 exposures of 120 s (Fig.\,\ref{findchart}). We used a random offset of 4\arcsec\ among images in order to perform background subtraction of the variable infrared sky. VLT/NaCo science images were reduced based on the standard tools provided by the ESO \textsc{eclipse} package \citep{devillard97}. The final image quality corresponds to a point-source average FWHM of 0\farcs38. In order to reduce as much as possible the effects of contamination due to nearby objects, relative aperture and point-spread function (PSF) photometry was obtained within narrow annuli (about 1--1.5 FWHM depending on the seeing conditions), while the background was evaluated close to the object under analysis. Absolute photometry was derived by analysis of the best seeing frames. The $K_{\rm S}$ magnitude we infer from the final NaCo image for the counterpart of \src\ is $K_{\rm S}=15.03\pm0.06$ mag. This value is slightly different from others that can be found in various catalogues (for example, the IRSF/SIRIUS Magellanic Clouds Point Source Catalog by \citealt{kato07short} gives $K_{\rm S}=14.85\pm0.02$). In this respect, we note that infrared variability is not unusual among Be stars. In particular, very large variations (up to about two magnitudes) have been observed in several sources (see e.g. \citealt{reig05,rln07} for examples) and likey reflect changes in the circumstellar disk.
\\
\indent In order to register the \cxo\ coordinates of \src\ on our infrared images, we obtained the image astrometry by using the positions of 6 stars selected from the 2MASS catalogues and within the $\sim$$30\arcsec$$\times30\arcsec$ NaCo field of view of final images. The residual in the fit was of 0\farcs06 in each coordinate, converting to $\sim$$0\farcs1$ once the  2MASS absolute accuracy was included.\footnote{See http://www.ipac.caltech.edu/2mass/releases/allsky/doc.} The astrometrically-corrected \cxo\ position of \src\ is $\rm RA = 00^h50^m48\fs03$, $\rm Decl.=-73\degr18'18\farcs2$ (J2000.0, with a 1$\sigma$ uncertainty of $0\farcs1$ on each coordinate).

\section{Optical Observations}
The optical counterpart of \src\ was observed spectroscopically with the 1.5-m CTIO telescope of Cerro Tololo (Chile) equipped with the R-C spectrograph, which carries a 1274$\times$280 pixels Loral CCD. Two 1500-s spectroscopic frames were secured on 2012 October 25, with start times at 05:43 and 06:26 UT, respectively. Data were acquired using Grating \#13I and with a slit width of $1\farcs5$, giving a nominal spectral coverage between 3300 and 10500 \AA~and a dispersion of 5.7 \AA\ pixel$^{-1}$.

After cosmic-ray rejection, the spectra were reduced, background subtracted and optimally extracted \citep{horne86} using \textsc{iraf} \citep{tody93}.\footnote{\textsc{iraf} (Image Reduction and Analysis Facility) is available at http://iraf.noao.edu/} Wavelength calibration was performed using He--Ar lamps acquired before each spectroscopic exposure; the spectra were then flux-calibrated using the spectro-photometric standard Feige~110 \citep{hamuy92,hamuy94}. Finally, the two spectra were stacked together to increase the signal-to-noise ratio. The wavelength calibration uncertainty was $\sim$0.5 \AA; this was checked using the positions of background night sky lines.

\subsection{Optical classification}

\begin{figure*}
\resizebox{\hsize}{!}{\includegraphics[angle=-90]{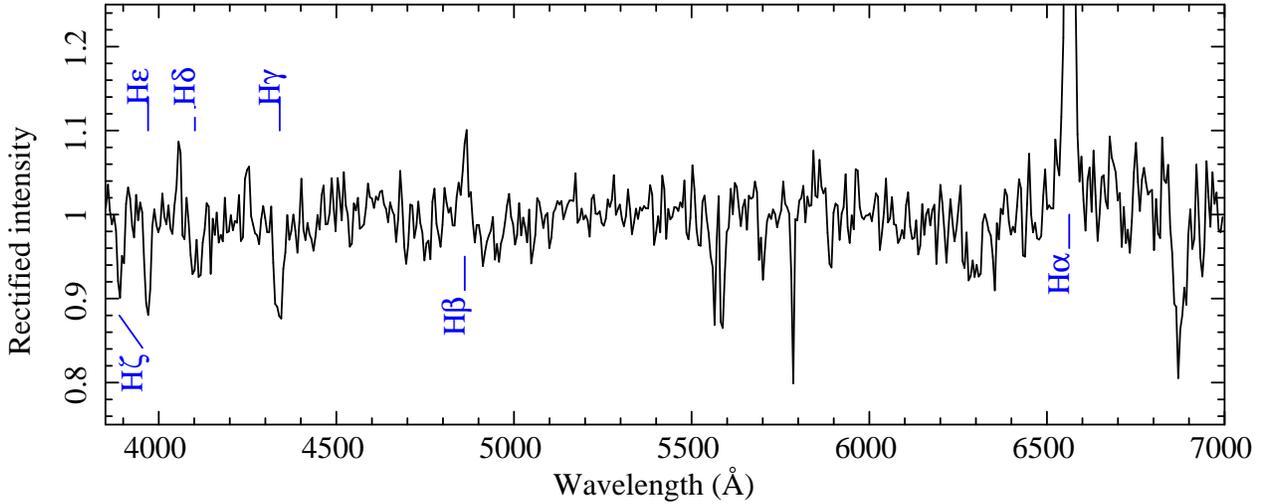}}
\caption{CTIO R-C spectrograph low-resolution (15 \AA; 3700--7000 \AA) rectified spectrum obtained for the optical counterpart of \src\ (effective exposure time of 3000 s). The main identified lines are marked with solid line and element identification.}
\label{fig:spectiao}
\end{figure*}

The optical spectrum is typical of early type OB stars with a strong blue continuum and shows the presence of only H$\alpha$ (equivalent width $EW\sim-31$ \AA) and H$\beta$ ($EW\sim-1.3$ \AA) in emission (H$\delta$ seems to be filled in, while H$\epsilon$ is in absorption) while H$\gamma$, H$\epsilon$ and H$\chi$ are in absorption (Fig.\,\ref{fig:spectiao}). However the low resolution and the relatively low signal to noise does not make it possible to use the usual spectral absorption features to classify the spectrum. We note the possible presence of O\textsc{ii} around 4061--63 \AA\ (and a hint of O\textsc{ii} at 4253--54 \AA) which would suggest a late O star. The H$\alpha$ profile is single-peaked and compatible, within the uncertainties, with a Gaussian with FWHM of $\sim$7 \AA, which is consistent with the instrumental resolution.

In order to constrain the spectral classification we used the information related to the known distance modulus (distance of 61 kpc). Assuming typical colours of stars in the O5--B2 range and luminosity classes between V and I \citep{wegner94}, and comparing with the observed magnitudes and colours reported in the optical catalogues\footnote{See http://vizier.u-strasbg.fr/viz-bin/VizieR.} (we assumed $14.50<V<15.62$ and $14.47<B<15.35$), we derived an excess colour interval of $0.08<E(B-V)<0.35$. These values, convert in an absolute magnitude in the range from $-5.4$ to $-4.5$. The latter values correspond to the following spectral types and luminosity classes \citep{gray09}: O6--O9V, O9--B0IV, B0--B1III and B1--B2II. This finding together with the H$\alpha$ $EW$ suggest a late O or early B V--IV luminosity class star, though more evolved companions cannot be ruled out by the present data. Finally, we note that a slitless spectroscopic study of the SMC stars carried out recently classified the optical counterpart of \src\ as a B0 star, further pushing towards a V--III luminosity class object \citep*{martayan10}. A medium/high resolution spectroscopic study is needed in order to confirm our conclusions.

\section{Discussion}
The observed properties of \src\ indicate a massive X-ray binary, with a neutron star orbiting an early type companion, most likely a main sequence O or B-type star, naturally suggesting a Be/X-ray binary (Be/XRB) located in the SMC. This is not unusual, since SMC hosts an overabundance of Be/XRBs compared to the Milky Way and Be/XRBs are the dominant population of high mass X-ray binaries in the SMC \citep{coe05}.

The X-ray pulsar is therefore accreting matter from the circumstellar decretion disc of the Be donor. In such systems, the transient X-ray emission can be due to either  enhanced wind accretion in a wide and highly eccentric orbit when the neutron star approaches the periastron (periodic type I outbursts, reaching 10$^{36}$~erg~s$^{-1}$ at the peak) lasting for a fraction of 0.2--0.3 of the orbital period, or to an enhanced mass loss from the disruption of the Be disc, producing more luminous X-ray emission ($10^{37}$--$10^{38}$~erg~s$^{-1}$) which can last for weeks or months (aperiodic type II, or `giant',  outbursts; \citealt{negueruela98}).

The discovered pulsar spin period of 292.7~s, assuming the observed correlation of the spin versus orbital period in Be/XRBs in the Corbet diagram \citep{corbet86}, implies an orbital period roughly in the range 50--300~days. \src\ reached an outburst peak of $\sim$$10^{36}$~erg~s$^{-1}$ during a transient activity lasting for about 20 days, with a luminosity in excess of 10$^{35}$~erg~s$^{-1}$. A type II outburst in a Be/XRB cannot be ruled out in this source, also considering that the orbital geometry is not known (we note for example that type I outburst are more likely to occur in systems whith high eccentricity; \citealt{okazaki01,reig07}). However, given the expected range for the orbital period and the duration of the outburst, as well as its low peak X-ray luminosity, we tend to prefer an interpretation of the transient activity of \src\ as a type I outburst. In this hypothesis, we can compare the observed X-ray light curve with what  expected in terms of Bondi--Hoyle accretion of matter from the equatorial disc-wind of the Be companion (see equations in \citealt{wvk89}). This decretion disc is assumed to display a power-law density distribution, as follows:
\begin{equation}
\rho_{\mathrm{wind}}(r) = \rho_{0} \left( \frac{r}{R_{*}} \right)^{-n}
\end{equation}
where $\rho_{0}$ is the density at the surface of the star ($\sim$10$^{-11}$~g~cm$^{-3}$) and $R_{*}$ is the stellar radius. This translates into the velocity law:
\begin{equation}
v_{\mathrm{wind}}(r) = v_{0} \left( \frac{r}{R_{*}} \right)^{n-2}.
\end{equation}
The outflow displays also a rotational component of the velocity, described as: 
\begin{equation}
v_{{\rm rot, w}}(r) = v_{{\rm rot,0}} \left( \frac{r}{R_{*}}
\right)^{-\alpha}
\end{equation}
where $\alpha=0.5$ implies a Keplerian decretion disc.
The relative velocity between the matter in the disc and the neutron star is:
\begin{equation}
v^{2}_{{\rm rel}} = \left( v_{\mathrm{wind}}(r) - V_{{\rm r}}\right)^{2} + 
 \left( v_{{\rm rot,w}}(r) - V_{{\rm \theta}}\right)^{2} + c_{\mathrm{S}}^2
\end{equation}
where $c_{\mathrm{S}}$ is the speed of sound, and  $V_{{\rm r}}$ and  $V_{{\rm \theta}}$ are the radial and tangential components of the neutron-star velocity along the orbit, respectively.

\begin{figure}
\centering
\resizebox{\hsize}{!}{\includegraphics[angle=0]{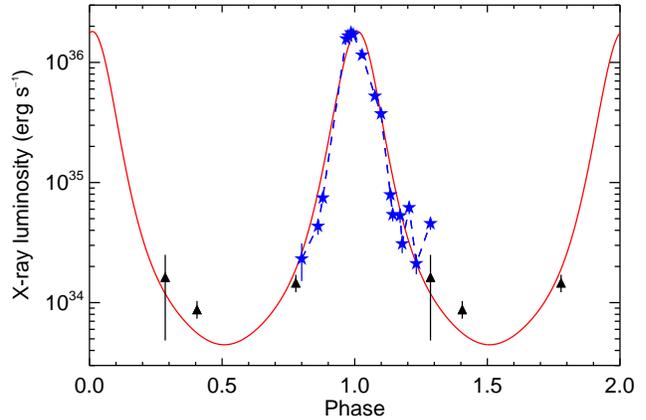}}
\caption{\label{fig:calc} Comparison between the observed light curve (here, the 1--10 keV X-ray luminosity, $L_{\rm X}$, has been multiplied by a factor 1.5 to better approximate the whole X-ray luminosity) and the X-ray emission (solid line) calculated from Bondi--Hoyle accretion assuming a Be outflowing disc wind with the parameters listed in Table~\ref{tab:wind}, together with an orbital period, $P_{\rm orb}$, of 100~days, an eccentricity $e=0.44$, and an exponent, $n$, of the density distribution, of 3.1. Stars indicate the consecutive observations during the 2010 outburst, while triangles represent the three observations performed in 2002, 2003 and 2006, folded on the same orbital period.}
\end{figure}

All the above parameters are completely unknown in  \src.  So, the best we can do to compare the observed and the calculated X-ray light curve is to fix most of these parameters to reasonable values, as reported in Table~\ref{tab:wind}, except the orbital period $P_{\rm orb}$, the orbital eccentricity $e$, and the exponent of the density distribution, $n$, which are varied in order to reproduce the observed light curve. We caution that there is a degeneracy of these parameters, implying that very different values for this triad can reproduce the observed X-ray light curve equally well. For this reason, we will only give an example of this comparison in Fig.~\ref{fig:calc}, where the observed light curve is  compared  with the calculated one, assuming $P_{\rm orb}=100$~days, $e=0.44$ and $n= 3.1$. Since the observed X-ray luminosities are in the 1--10~keV energy band, they represent a lower limit to the X-ray (bolometric) luminosity which results from the Bondi--Hoyle accretion model. For this reason, in Fig.~\ref{fig:calc} we multiplied the 1--10~keV light curve by 1.5.
\begin{table}
\centering \caption{Assumed parameters for the calculated X-ray light curve.} \label{tab:wind}
\begin{tabular}{@{}lc}
\hline
Parameter  & Value       \\
       &           \\
\hline
Companion mass, $M_{*}$   &    17 $M_\odot$  \\
Companion radius, $R_{*}$  &   10 $R_\odot$    \\
$\rho_{0}$         &    10$^{-11}$~g~cm$^{-3}$   \\
$v_{0}$           &   30~km~s$^{-1}$    \\
$v_{{\rm rot,0}}$  &  300~km~s$^{-1}$      \\
$\alpha$ &     0.5   \\
Wind temperature, $T$ &    $2.5\times10^{4}$~K     \\
\hline
\end{tabular}
\end{table}

Note that the three observations performed outside the outburst in 2002, 2003 and 2006 (triangles in Fig.~\ref{fig:calc}) represent a further constraint on the allowed parameters. In fact, as an example, we note that to model the outburst light curve, an equally good choice for the three parameters would be  $P_{\rm orb}=50$~days, $e=0.28$ and $n= 3.3$. However, when also the three 2002--2006 data points, folded on a period of 50~days, are compared with the theoretical light curve, they fall in the region occupied by the outburst peak. This excludes this specific combination of $P_{\rm orb}$, $e$ and $n$, at least assuming that there is an outburst at every periastron passage. 

We stress that although the orbital period of 100~days lies in the expected range  for a Be/XRB pulsar with a spin period of 292.7~s, it {\it should not} be considered as the true orbital period of \src.  This comparison only meant to demonstrate that the source outburst can be well explained as a type I outburst in a Be/XRBs, within a simple direct accretion model and assuming standard values for the system (orbital and wind) parameters.

\section*{Acknowledgments} 
This research is based on data and software provided by the CXC (operated for NASA by SAO) and on observations carried out with the ESO/VLT 8.2-m Unit 4 (Yepun) telescope and the NOAO/CTIO SMARTS 1.5-m telescope. We thank Rodrigo Hern\'andez for Service Mode observations at the 1.5-m CTIO telescope and Fred Walter for coordinating them. We are grateful to the referee, Malcolm Coe, for his valuable and constructive comments. \textsc{iraf} is made available to the astronomical community by the National Optical Astronomy Observatories, which are operated by AURA, Inc., under contract with the U.S. National Science Foundation.

\bibliographystyle{mn2e}
\bibliography{biblio}

\bsp

\label{lastpage}

\end{document}